\documentstyle[aps]{revtex}
\draft
\def \D {\mbox{D}}
\def \curl {\mbox{curl}\,}

\begin{document}
\twocolumn[\hsize\textwidth\columnwidth\hsize\csname
@twocolumnfalse\endcsname

\title{Magnetized gravitational waves}

\author{Roy Maartens, Christos G. Tsagas and Carlo Ungarelli}
\address{~}
\address{Relativity and Cosmology Group, School of
Computer Science and Mathematics,\\ Portsmouth University,
Portsmouth~PO1~2EG, Britain}

\maketitle

\begin{abstract}

We investigate the influence of cosmic magnetic fields on
gravitational wave perturbations, and find exact solutions on
large scales. We show that a large-scale magnetic field can
generate large-scale non-decaying gravitational waves. In the
general case where gravitational waves are generated by other
mechanisms, a large-scale magnetic field introduces a new decaying
tensor mode and modifies the non-decaying mode. The direct effect
of the magnetic field is to damp the gravitational waves, while an
indirect magneto-curvature effect can either damp or boost the
waves. A magnetic field also leads to a breaking of statistical
isotropy, and the magnetic imprint on the tensor spectrum in
principle provides a means of detecting a primordial field.
\\\\
PACS number(s): 98.80.Hw, 04.40.Nr, 95.30.Qd, 98.62.En
\end{abstract}\vskip2pc]

\section{Introduction}
%%%%%%%%%%%%%%%%%%%%%%

Magnetic fields seem to be everywhere that we can look in the
universe, from our own sun out to high-redshift Lyman-$\alpha$
systems. The fields we observe (based on synchrotron radiation and
Faraday rotation) in galaxies and clusters have been amplified by
gravitational collapse and possibly also by dynamo mechanisms.
They are either primordial, i.e. originating in the early universe
and already present at the onset of structure formation, or they
are protogalactic, i.e. generated by battery mechanisms during the
initial stages of structure formation. One way to distinguish
these possibilities would be to detect or rule out the presence of
fields coherent on cosmological scales during recombination via
their imprint on the cosmic microwave background (CMB) radiation.
Considerable work has been done to investigate the nature of the
magnetic imprint on the CMB (see~\cite{mag} for recent reviews and
further references).

We show here that cosmological magnetic fields also leave a
characteristic imprint on the cosmological gravitational wave
background. Direct detection of this background, especially on
large scales, is not likely for a considerable time (see,
e.g.,~\cite{m}), so that our results do not provide a practical
means for detecting or limiting a large-scale magnetic field.
However, these results are a necessary first step in a theoretical
understanding that can be developed in anticipation of
gravitational wave detection. Furthermore, our results suggest a
similar investigation of the magnetic imprint on gravitational
waves generated by compact objects. This astrophysical problem is
more complex than the cosmological problem that we discuss, but it
is likely to lead to larger effects with stronger prospects for
observational detection.

The effects of gravitational waves on electromagnetic fields, in
particular, the generation of electromagnetic pulses by
gravitational waves, has been previously considered, including the
implications for gravitational wave detection~\cite{bg}
(see~\cite{mbd} for recent results and further references). Here
we consider a different and new effect, i.e. the influence of a
large-scale magnetic field on gravitational waves, including the
capacity of the field to generate gravitational waves. This latter
possibility has been previously investigated on small scales
in~\cite{dfk}, where tangled magnetic fields during recombination
were shown to generate gravitational wave perturbations, which in
turn contribute to CMB anisotropies, thus providing a way of
limiting the strength of the magnetic field.

We consider in a general cosmological context the problem of how
magnetic fields affect gravitational waves, focusing on
large-scale fields. A large-scale homogeneous magnetic field is
strongly limited by large-angle CMB anisotropies~\cite{bfs}:
$B_{\rm now} \lesssim 10^{-9}\,$G. Although the maximal energy
density supported by the field is a very small fraction of the
total cosmic energy density, this is also the case for
gravitational waves, so that in principle, as we show, the
magnetic effect on these waves need not be negligible. We find
exact analytic solutions on large scales. These solutions show
that the direct effect of the magnetic field is to damp the
gravitational waves, while an indirect effect due to the coupling
between the field and the spatial curvature can either damp or
boost the waves. Qualitatively, the presence of a large-scale
cosmic magnetic field is signalled by a breaking of global
statistical isotropy in the gravitational wave spectrum, which is
in principle detectable (anisotropy detection is discussed in
general in~\cite{ao}).

\section{Perturbation equations}
%%%%%%%%%%%%%%%%%%%%%%%%%%%%%%%%

The background is a spatially flat Friedmann spacetime with a
non-magnetized perfect fluid, representing non-interacting
baryons, radiation and cold dark matter with the same 4-velocity
$u^a$, and Hubble rate $H=\dot a/a$. The background equations are
\begin{eqnarray}
\dot{\rho}&=&-3H(\rho+p)\,,  \label{bce}\\ 3H^2&=&\rho+ \Lambda\,,
\label{bFe}
\end{eqnarray}
where $\rho$ and $p$ are the total energy density and pressure.
The perturbed universe is permeated by a weak large-scale magnetic
field $B_a$, whose energy density $\rho_{\rm mag}=
{\textstyle{1\over2}}B^2$ is a first-order quantity (so that $B_a$
is ``half-order"~\cite{dfk}). The magnetic field also has
first-order isotropic pressure ${\textstyle{1\over6}}B^2$ and
tracefree anisotropic stress
\begin{equation}\label{as}
\pi_{ab}=-B_{\langle a}B_{b\rangle}\equiv - [h_{(a}{}^ch_{b)}{}^d-
{\textstyle{1\over 3}}h^{cd}h_{ab}]B_cB_d\,,
\end{equation}
where $h_{ab}=g_{ab}+u_au_b$ projects into the comoving rest space
and the round brackets on indices denote symmetrization. The
baryonic fluid is treated as a perfect fluid of infinite
conductivity, with energy density $\rho_{\rm b}$ and isotropic
pressure $p_{\rm b}$. High conductivity ensures that any electric
fields that might have been present dissipate quickly. For tensor
perturbations, which do not excite relative velocity
perturbations, $u^a$ is the 4-velocity of baryons, radiation
($p_{\rm r}={1\over3}\rho_{\rm r}$) and collisionless cold dark
matter.

Both the isotropic and anisotropic magnetic effects are of the
same perturbative order as gravitational waves. The covariant
Maxwell equations~\cite{mb} imply the induction equation, whose
nonlinear form is~\cite{TB1}
\begin{equation}
h_a{}^b\dot{B}_{b}=\left(\sigma_{ab}
+\varepsilon_{abc}\omega^c-{\textstyle{2\over3}}\Theta
h_{ab}\right)B^b\,,  \label{dotHa}
\end{equation}
where $\sigma_{ab}$, $\omega_a$ and $\Theta$ are respectively the
shear, vorticity and expansion. In the background, $\Theta=3H$ and
$\omega_a=0=\sigma_{ab}$. The induction equation implies the
linearized conservation equations~\cite{TB1,t2}
\begin{eqnarray}
\left(B^2\right)^{\displaystyle{\cdot}}&=& -4HB^2\,,
\label{dotH2}\\ \dot{\pi}_{ab}&=& -4H\pi_{ab}\,, \label{dotPi}
\end{eqnarray}
where the dot denotes $u^a\nabla_a$. It follows that
\begin{equation}\label{1}
\pi_{ab}=-B_0^2\left({a_0\over a}\right)^4n_{\langle
a}n_{b\rangle}\,,~~n_a={B_a\over B}\,,~~\dot{n}^a=0\,.
\end{equation}
The unit magnetic direction vector $n^a$ is parallel-propagated
along each observer world-line to first order. Maxwell's equations
also imply the constraint~\cite{TM2}
\begin{equation}
{\rm D}^b\pi_{ab}=\varepsilon_{abc}B^b{\curl}B^c-
{\textstyle{1\over6}}{\rm D}_aB^2\,,  \label{Picon}
\end{equation}
where $\D_a$ is the projected covariant derivative:
$\D_aS^{b\cdots}{}{}_{c\cdots}=
(\nabla_aS^{b\cdots}{}{}_{c\cdots})_\perp$, which leads to
covariant spatial curl operators: $\curl
S_a=\varepsilon_{abc}\D^bS^c$, $~\curl S_{ab}=
\varepsilon_{cd(a}{\rm D}^cS^d{}_{b)}$.

The 4-acceleration $A_a$ is determined by Euler's
equation~\cite{TB1,t2}
\begin{eqnarray}
\left(\rho_{\rm b}+p_{\rm b}\right)A_a=-{\rm D}_ap_{\rm b}
 +\varepsilon_{abc}B^b{\curl}B^c\,.
\label{mdcl}
\end{eqnarray}
For tensor perturbations, the 4-acceleration should vanish at
first order. The non-magnetized condition for transverse traceless
(pure tensor) modes is~\cite{DBE,C}
\begin{equation}
\omega_a=0={\rm D}_a\rho=\D_ap\,. \label{nmgw}
\end{equation}
When there is no magnetic field, this ensures that
$A_a=0=\D_a\Theta$ and that all the traceless tensors, such as
$\sigma_{ab}$, are transverse. In the magnetized case, two
additional constraints need to be imposed:
\begin{equation}
{\rm D}_aB^2=0=\varepsilon_{abc}B^b{\curl}B^c\,,  \label{mgw}
\end{equation}
i.e., {\em the magnetic energy density is homogeneous and
the field is force-free,} to first
order. Equations~(\ref{nmgw}) and (\ref{mgw})
together imply
\begin{equation}
A_a=0=\D_a\Theta\,,
\end{equation}
and that all traceless tensors are transverse. (The expansion
gradient is seen to vanish from the propagation equation for
$\D_a\rho$~\cite{TB1}.) The magnetic field is felt only through
its transverse traceless anisotropic stress $\pi_{ab}$.

Gravitational radiation is covariantly described by the electric
($E_{ab}=E_{\langle ab\rangle}$) and magnetic ($H_{ab}=H_{\langle
ab\rangle}$) parts of the Weyl tensor, which support different
polarization states~\cite{C}, and which obey equations remarkably
analogous to Maxwell's~\cite{mb,C,DBE}:
\begin{eqnarray}
\dot{E}_{ ab}&=& -3H E_{ab}+ {\textstyle{3\over2}}H\pi_{ab}-
{\textstyle{1\over2}}(\rho+p)\sigma_{ab}+ \curl H_{ab}\,,
\label{pldotE}\\ \dot{H}_{ab}&=&-3H  H_{ab}- \curl  E_{ab}+
{\textstyle{1\over2}}\curl \pi_{ab}\,, \label{pldotH}\\
\dot{\sigma}_{ab}&=& -2H\sigma_{ab}- E_{ab}+
{\textstyle{1\over2}}\pi_{ab}\,,  \label{pldotsigma}\\
H_{ab}&=&\curl \sigma_{ab}\,, \label{plHab}
\end{eqnarray}
where
\begin{equation}
\D^bE_{ab}=\D^bH_{ab}=\D^b\sigma_{ab}=\D^b\pi_{ab}=0\,.
\end{equation}
In the magnetized case, $\pi_{ab}$ is given by Eq.~(\ref{1}).

The direct relation between the shear and the magnetic Weyl tensor
means that we can describe gravity wave evolution via $E_{ab}$ and
$\sigma_{ab}$ alone. Equations (\ref{pldotE})--(\ref{pldotsigma})
show how the magnetic field is a source of shear and gravitational
wave perturbations via its anisotropic pressure.
Equation~(\ref{plHab}) allows us to decouple Eq.~(\ref{pldotH})
from the system, which reduces to Eq.~(\ref{pldotsigma}) and
\begin{eqnarray}
\dot{E}_{ ab}= -3H  E_{ab}+ {\textstyle{3\over2}}H\pi_{ab}-
{\textstyle{1\over2}}(\rho+p)\sigma_{ab}- {\rm D}^2\sigma_{ab}\,,
\label{1pldotE}
\end{eqnarray}
on using the linearized identity $\curl\curl S_{ab}=-{\rm
D}^2S_{ab}$. It follows via the role of $\pi_{ab}$ in these
equations that there is a directional effect at all scales on
gravitational waves due to the magnetic field. We will further
discuss this effect on large scales below. Note also that to first
order, the evolution of the magnetic field is independent of any
gravity wave effects. The field affects gravity waves, but there
is no backreaction on the field.

We can further reduce the system of equations to a single
covariant wave equation in $\sigma_{ab}$:
\begin{eqnarray}
&&{\rm D}^2\sigma_{ab}-\ddot{\sigma}_{ab} = 5H\dot{\sigma}_{ab}+
\left[{\textstyle{1\over2}}(\rho-3p)+2\Lambda\right] \sigma_{ab}
\nonumber\\&&~~{} +2B_0^2\left({a_0\over a}\right)^4Hn_{\langle
a}n_{b\rangle} \,, \label{lddotsigma}
\end{eqnarray}
on using Eqs.~(\ref{bce}), (\ref{bFe}), (\ref{1}), (\ref{pldotE})
and (\ref{plHab}). The solution of this wave equation then gives
the gravito-electromagnetic tensors via Eqs.~(\ref{pldotsigma})
and (\ref{plHab}). In practice, it is better to solve the first
order coupled system for $E_{ab}$ and $\sigma_{ab}$ than the
single second order wave equation.

The covariant Maxwell-Weyl approach to gravitational wave
perturbations may be related to the metric-based approach (see
also~\cite{DBE,C,tm}). The transverse traceless metric
perturbation $f_{ij}$ is defined by
\begin{equation}
ds^2=-dt^2+a^2[\delta_{ij}+f_{ij}]dx^idx^j\,.
\end{equation}
In these coordinates, it follows that
$E_{0a}=0$ and~\cite{g}
\begin{equation}
E_{ij}=-{\textstyle{1\over4}}\left[\ddot{f}_{ij}+H\dot{f}_{ij}
\right]\,,
\end{equation}
on large scales. The wave equation for $f_{ij}$ then shows that,
neglecting anisotropic stresses, $E_{ij}={1\over2}H\dot{f}_{ij}$.
Thus the energy density in gravitational waves, $\rho_{\rm
gw}={1\over4}\dot{f}_{ij}\dot{f}^{ij}$, is given in the absence of
a magnetic field in Maxwell-Weyl form as
\begin{equation}\label{en}
\rho_{\rm gw}={1\over H^2}E_{ab}E^{ab}\,
\end{equation}
on large scales. Whether or not there is a magnetic field, the
dimensionless contribution to the power spectrum on large scales
is determined by $\lambda^4E_{ab}E^{ab}$, where $\lambda$ is the
wavelength~\cite{C}. Thus a dimensionless measure of amplitude on
large scales is given by
\begin{equation}\label{obs}
\Gamma^2=a^4E_{ab}E^{ab}\,.
\end{equation}

\section{Large-scale solutions}
%%%%%%%%%%%%%%%%%%%%%%%%%%%%%%%

On superhorizon scales we can neglect the Laplacian term in
Eq.~(\ref{1pldotE}). In the radiation era, Eqs.~(\ref{pldotsigma})
and (\ref{1pldotE}) then have the solution
\begin{eqnarray}
E^{ab}&=&C^{ab}_{+}\tau^{-1}+ C^{ab}_{-}\tau^{-5/2}-
{\textstyle{1\over2}} n^{\langle a}n^{b\rangle}{B_0^2}\tau^{-2}\,,
\label{rE}\\ H_0{\sigma^{ab}}
&=&2C^{ab}_{+}-C^{ab}_{-}\tau^{-3/2}-
{\textstyle{1\over2}} n^{\langle a}n^{b\rangle} B_0^2\tau^{-1}\,,
\label{rsigma}
\end{eqnarray}
where $\dot{C}^{ab}_{\pm}=0$ and $\tau\equiv t/t_0$. In the matter
era:
\begin{eqnarray}
E^{ab}&=&C^{ab}_{+}\tau^{-4/3} +C^{ab}_{-}
\tau^{-3}-{\textstyle{3\over4}} n^{\langle
a}n^{b\rangle}{B_0^2}\tau^{-8/3}\,, \label{dE}\\H_0{\sigma^{ab}}
&=&C^{ab}_{+}\tau^{-1/3}- {\textstyle{3\over2}}
C^{ab}_{-}\tau^{-2} -{\textstyle{1\over3}} n^{\langle
a}n^{b\rangle} {B_0^2} \tau^{-5/3} \,. \label{dsigma}
\end{eqnarray}

The magnetic field introduces a new mode of gravitational wave
perturbations on large scales, which decays less rapidly than the
standard, non-magnetized decaying mode, and it modifies the
standard non-decaying mode (of $a^2E_{ab}$), i.e., magnetic terms
enter $C_{+}^{ab}$ (see below). The directional influence of the
large-scale magnetic field on gravitational waves means that the
field breaks the statistical isotropy of the large-scale tensor
spectrum, as further discussed below.

The new magnetized modes mean that {\em if there are no
gravitational wave perturbations initially present, then these can
be generated by a large-scale magnetic field}. (The generation of
tensor perturbations by small-scale tangled magnetic fields is
investigated in Ref.~\cite{dfk}.) This could happen if a
large-scale magnetic field is created at time $t_0$. Large-scale
magnetogenesis can occur for example in the recombination
era~\cite{h} or the preheating era after inflation (see~\cite{bp}
and references cited therein). Since
$E_{ab}(t_0)=0=\dot{E}_{ab}(t_0)$, it follows from Eq.~(\ref{rE})
that
\begin{equation}\label{gen}
E_{ab}=\left[{\textstyle{5\over6}}\tau^{-1}-
{\textstyle{1\over3}}\tau^{-5/2} -{\textstyle{1\over2}}\tau^{-2}
\right]B_0^2n_{\langle a}n_{b\rangle}\,,
\end{equation}
if $t_0$ is in the radiation era, with a similar result if $t_0$
is in a matter-dominated era. These `purely magnetic'
gravitational waves have a non-decaying mode,
\begin{equation}
\Gamma={5\sqrt{6}\over18}\,a_0^2B_0^2\,,
\end{equation}
and they satisfy $E_{ab}\propto\pi_{ab}$, so that in some sense
they are maximally anisotropic (see below). The generation of
tensor perturbations via magnetogenesis may be compared with the
magnetic generation of density perturbations, which produces a
growing mode~\cite{kor,t2}.

In the general case, tensor perturbations are generated by other
mechanisms and then influenced by the magnetic field. The
dimensionless measure of amplitude, $\Gamma$, defined in
Eq.~(\ref{obs}), is seen from Eqs.~(\ref{rE})--(\ref{dsigma}) to
be constant at late times ($t\gg t_0$), i.e., when we neglect the
decaying modes. Explicitly, we find the following magnetic
influence on the non-decaying mode of large-scale gravitational
waves.\\ In the radiation era,
\begin{eqnarray}
\Gamma^2&=&{{1\over9}} \left[4\Gamma_0^2+4a_0^2H_0^2\Sigma_0^2-8
a_0H_0 {\cal X}_0\right] \nonumber\\&&~{}
+{{2\over27}}a_0^2B_0^2\left(a_0^2B_0^2-6{\cal E}_0 +6a_0H_0{\cal
S}_0 \right)\,, \label{rE2}
\end{eqnarray}
where we expressed $C_+^{ab}$ in terms of the dimensionless
physical scalars,
\begin{eqnarray}
&&\Sigma^2=a^2\sigma_{ab}\sigma^{ab}\,,~~{\cal
S}=a\sigma_{ab}n^an^b\,,\\ && {\cal E}=a^2E_{ab}n^an^b\,,~~{\cal
X}=a^3E_{ab}\sigma^{ab}\,.
\end{eqnarray}
In the matter era,
\begin{eqnarray}
\Gamma^2&=&{{9\over25}} \left[\Gamma_0^2+a_0^2H_0^2\Sigma_0^2-2
a_0H_0 {\cal X}_0\right] \nonumber\\&&~{}
+{{3\over50}}a_0^2B_0^2\left(a_0^2B_0^2-6{\cal E}_0 +6a_0H_0{\cal
S}_0 \right)\,. \label{dE2}
\end{eqnarray}

The effect of the magnetic field is more clearly brought out if we
use the Gauss-Codazzi equation~\cite{TB1}
\begin{equation}
R^*_{\langle ab\rangle}=E_{ab}-H\sigma_{ab}+
{\textstyle{1\over2}}\pi_{ab}\,,
\end{equation}
where $R^*_{ab}$ is the Ricci tensor of the spatial hypersurfaces
orthogonal to $u^a$. This equation implies
\begin{equation}\label{gc}
{\cal R}={\cal E}-aH{\cal S}- {\textstyle{1\over3}}a^2B^2\,,
\end{equation}
where
\begin{equation}
{\cal R}=a^2R^*_{\langle ab\rangle}n^an^b
\end{equation}
is a dimensionless curvature scalar, giving the anisotropic
3-Ricci curvature along the magnetic field, due to the
gravitational waves. Using Eq.~(\ref{gc}), we can rewrite
Eqs.~(\ref{rE2}) and (\ref{dE2}):
\begin{eqnarray}
\Gamma^2&=&{{1\over9}} \left[4\Gamma_0^2+4a_0^2H_0^2\Sigma_0^2-8
a_0H_0 {\cal X}_0\right] \nonumber\\&&~{}
-{{2\over27}}a_0^2B_0^2\left(a_0^2B_0^2+6{\cal R}_0 \right)\,,
\label{rE3}
\end{eqnarray}
in the radiation era, and
\begin{eqnarray}
\Gamma^2&=&{{9\over25}} \left[\Gamma_0^2+a_0^2H_0^2\Sigma_0^2-2
a_0H_0 {\cal X}_0\right] \nonumber\\&&~{}
-{{3\over50}}a_0^2B_0^2\left(a_0^2B_0^2+6{\cal R}_0\right)\,,
\label{dE3}
\end{eqnarray}
in the matter era.

These equations show that there are two aspects to the magnetic
effect: a `pure' magnetic effect (neglecting curvature), and a
magneto-curvature effect. The pure effect, proportional to
$-B_0^4$, serves to damp the amplitude relative to the
non-magnetized case. It is due to the tension of the magnetic
field lines, which means that the magnetic field resists the
distortions induced by a gravitational wave~\cite{t}. The
magneto-curvature effect, proportional to $-B_0^2{\cal R}_0$,
arises from the geometric coupling between curvature and the field
in general relativity, due to the vector nature of the
field~\cite{TB1,t2,TM2,t}. It depends on the sign of ${\cal R}_0$,
i.e., on the sign of the anisotropic 3-Ricci curvature component
along the magnetic direction at time $t_0$. If ${\cal R}_0>0$,
then the damping is reinforced; otherwise, the magneto-curvature
effect enhances the amplitude, acting opposite to the pure
magnetic effect.

\section{Magnetic imprint on gravitational waves}
%%%%%%%%%%%%%%%%%%%%%%%%%%%%%%%%%%%%%%%%%%%%%%%%%

In the above solutions, the terms in round brackets give the
magnetic correction to the non-magnetized model. A rough estimate
of the relative magnitude of the magnetic imprint on the
large-scale gravitational wave spectrum is therefore
\begin{equation}
\alpha\equiv{\Gamma^2\over \Gamma^2|_{B=0}}\sim
{a_0^4B_0^4\over\Gamma_0^2}\,.
\end{equation}
On using Eqs.~(\ref{bFe}), (\ref{en}) and (\ref{obs}), we find
\begin{equation}\label{mi}
\alpha \sim \left({\rho_{\rm mag}^2\over H^2\rho_{\rm
gw}}\right)_0 \sim {(\rho_{\rm mag}/\rho_{\rm r})_0^2 \over
(\rho_{\rm gw}/\rho_{\rm r})_0}\left({\rho_{\rm r}\over \rho}
\right)_0\,.
\end{equation}
Now $\rho_{\rm mag}/\rho_{\rm r}$ is constant to first order, and
large-angle CMB anisotropies at the $10^{-5}$ level imply the
rough limit $(\rho_{\rm mag}/\rho_{\rm r})_{\rm dec}< 10^{-5}$.
The CMB quadrupole also places an upper limit on large-scale
tensor perturbations~\cite{l}: $(\Omega_{\rm gw}/ \Omega_{\rm r}
)_{\rm now}<10^{-10}$. Clearly $\alpha$ is sensitive to the tensor
contribution to CMB anisotropies: the smaller this contribution,
the higher $\alpha$ is. However, this increase in $\alpha$ comes
at the cost of greater difficulty in detecting the gravitational
wave background. In the most favourable case for detection, when
the tensor contribution to the CMB quadrupole is of the same order
as the scalar contribution and the magnetic contribution to the
CMB is maximal, then, with $t_0=t_{\rm dec}$, it follows that
$\alpha\lesssim (\rho_{\rm r}/ \rho)_{\rm dec}\sim 10^{-1} $.

Thus in principle, observations of the large-scale gravitational
wave background can provide quantitative limits on a possible
large-scale magnetic field, although in practice, detection of
such a background presents great difficulties (see~\cite{m} and
references therein). Thus the CMB remains the best way to limit a
magnetic field, on all scales~\cite{mag,dfk,bfs,sf,kkmd}.

Qualitatively, {\em the presence of a magnetic field is signalled
by the breaking of global statistical isotropy in the tensor
spectrum.} This may be compared with a similar situation regarding
the effect of a large-scale magnetic field on CMB polarization
anisotropy, where a magnetic presence is signalled by a
correlation between B-polarization and temperature anisotropies,
which is impossible if statistical isotropy holds~\cite{sf}. A
possible measure of global statistical anisotropy is the average
over all directions $n^a$ of the quantity
\begin{equation}\label{xi}
\xi={\Gamma^2-3{\cal E}^2\over \Gamma^2}\,.
\end{equation}
In the absence of a magnetic field, statistical isotropy will give
$\langle \xi\rangle_{\bf n}=0$, since
\begin{equation}
\langle (E_{ab}n^an^b)^2\rangle_{\bf
n}={\textstyle{1\over3}}E_{ab}E^{ab}\,.
\end{equation}
A measurable magnetic field will produce a non-zero $\xi$ for the
magnetic direction $n^a$. In the case of gravitational waves
generated by a magnetic field, Eqs.~(\ref{gen}) and (\ref{xi})
show that $|\xi|=1$.

Although we were only able to find exact solutions on large
scales, the small-scale solutions should share the same basic
features. Finally, we point out that if the same qualitative
physical effects of magnetic fields on gravitational waves occur
in the context of compact objects, then since the gravitational
and magnetic fields can be very strong, this could have more
immediate and important observational implications. This is a
topic currently under investigation.

\[ \]
{\bf Acknowledgments:} We would like to thank Anthony Challinor,
Ruth Durrer and Philippos Papadopoulos for helpful discussions and
comments. CGT is supported by PPARC.

\end{document}